# Transdisciplinary collaborations for advancing sustainable and resilient agricultural systems

Vesna Bacheva[1,2,3,§], Imani Madison[4,§], Mathew Baldwin[5], Mark Beilstein[6], Douglas F. Call[7], Jessica A. Deaver[7], Kirill Efimenko[8], Jan Genzer[8], Khara Grieger[9], April Z. Gu[5], Mehmet Mert Ilman[10,11], Jen Liu[12], Sijin Li[1], Brooke K. Mayer[13], Anand Kumar Mishra[10], Juan Claudio Nino[14], Gloire Rubambiza[12], Phoebe Sengers[12], Robert Shepherd[10], Jesse Woodson[6], Hakim Weatherspoon[12], Margaret Frank[2,*], Jacob Jones[15,*], Rosangela Sozzani[4,*], Abraham Stroock[1,3,*]

[§]These authors contributed equally
*Corresponding authors

[1]Smith School of Chemical and Biomolecular Engineering, Cornell University, Ithaca 14853, NY, USA
[2]School of Integrative Plant Science, Cornell University, Ithaca, NY 14853, USA
[3]Kavli Institute at Cornell for Nanoscale Science, Cornell University, Ithaca, NY 14853, USA
[4]Department of Plant and Microbial Biology and NC Plant Sciences Initiative, North Carolina State University, Raleigh, NC 27695, USA
[5]Department of Civil and Environmental Engineering, Cornell University, Ithaca, New York 14853, USA
[6]School of Plant Sciences, University of Arizona, Tucson, AZ 85721, USA
[7]Department of Civil, Construction, and Environmental Engineering & Science and Technologies for Phosphorus Sustainability (STEPS) Center, North Carolina State University, Raleigh, NC, 27695, USA
[8]Department of Chemical & Biomolecular Engineering, North Carolina State University, Raleigh, NC, 27695, USA
[9]Department of Applied Ecology and North Carolina Plant Sciences Initiative and Science and Technologies for Phosphorus Sustainability (STEPS) Center, North Carolina State University, Raleigh, NC, 27695, USA
[10]Department of Mechanical and Aerospace Engineering, Cornell University, Ithaca, NY 14850, USA
[11]Department of Mechanical Engineering, Hasan Ferdi Turgutlu Faculty of Technology, Manisa Celal Bayar University, Turgutlu 45400, Manisa, Turkey
[12]Department of Information Science, Cornell University, Ithaca, New York 14853, USA
[13]Department of Civil, Construction and Environmental Engineering and Science and Technologies for Phosphorus Sustainability (STEPS) Center, Marquette University, Milwaukee, WI 53233, USA
[14]Department of Materials Science and Engineering, University of Florida, Gainesville FL 32606, USA
[15]Department of Materials Science and Engineering and North Carolina Plant Sciences Initiative and Science and Technologies for Phosphorus Sustainability (STEPS) Center, North Carolina State University, Raleigh, NC, 27695, USA

**Abstract**

Feeding the growing human population sustainably amidst climate change is one of the most important challenges in the 21st century. Current practices often lead to the overuse of agronomic inputs, such as synthetic fertilizers and water, resulting in environmental contamination and diminishing returns on crop productivity. The complexity of agricultural systems, involving plant-environment interactions and human management, presents significant scientific and technical challenges for developing sustainable practices. Addressing these challenges necessitates transdisciplinary research, involving intense collaboration among fields such as plant science, engineering, computer science, and social sciences. Here, we present five case studies from two research centers demonstrating successful transdisciplinary approaches toward more sustainable water and fertilizer use. These case studies span multiple scales. Starting from whole-plant signaling, we explore how reporter plants can transform our understanding of plant communication and enable efficient application of water and fertilizers. We then show how new fertilizer technologies could increase the availability of phosphorus in the soil. To accelerate advancements in breeding new cultivars, we discuss robotic technologies for high-throughput plant screening in different environments at a population scale. At the ecosystem scale, we investigate phosphorus recovery from aquatic systems and methods to minimize phosphorus leaching. Finally, as agricultural outputs affect all people, we show how to integrate stakeholder perspectives and needs into the research. With these case studies, we hope to encourage the scientific community to adopt transdisciplinary research and promote cross-training among biologists, engineers, and social scientists to drive discovery and innovation in advancing sustainable agricultural systems.

**Introduction**

Feeding the growing human population in a sustainable way and in a changing climate is one of the most important challenges we face in the 21st century. Without substantial shifts in the agricultural practices on a global scale, we risk surpassing the planetary boundaries including freshwater use and biogeochemical flows of nitrogen (N) and phosphorus (P).[1] Currently, the best croplands are already under cultivation and much of the remaining land is unsuitable for farming. In addition, most of this cropland is rain fed, making crops vulnerable to unpredictable weather patterns. Other key inputs controlling plant productivity, such as synthetic fertilizers are often overused, with only a fraction effectively reaching the plants, while the remainder is lost to the environment.[2] Moreover, the rate of productivity improvement for many of our most important crops is slowing down.[3] Ensuring a healthy future for the people and the planet requires a deeper understanding of crop-based agriculture systems and their response to the environment.

Agriculture involves managing complex living systems embedded within complex ecological, societal, and economic systems that operate at multiple scales (Fig.1). At the farm level, management practices rely on predictive modeling, supported by measurements taken from soils and individual plants and the associated organisms that make up their microbiomes. Transitioning to a regional scale, effective coordination of shared resources such as land and water, alongside the development of policies, is essential to ensure sustainable production. On the global scale, concentrated efforts are necessary to address challenges including climate change and equitable food distribution, thereby ensuring food security worldwide. Changing agricultural practices would require an intense collaboration of all disciplines involved in these systems, including plant science, engineering, and computer science. Furthermore, given that agricultural outcomes impact everyone, this transdisciplinary effort must involve diverse fields of economics, policy, social sciences, and the engagement of various stakeholders including farmers and consumers.

The complexity of agricultural systems, including plants, their interaction with a local environment that spans from the atmosphere to below ground, and the human context in both management and consumption, presents numerous scientific and technical challenges in developing sustainable agricultural practices. The widespread use of synthetic fertilizers, primarily containing nitrogen and phosphorus, for decades has significantly boosted global food production capacity.[4] However, a major challenge arises from inefficient application of fertilizers, which contaminates the environment via leaching, particularly in water bodies where it damages aquatic ecosystems.[5] For example, crops utilize only around one-third of the nitrogen supplied through fertilization, with the remainder lost through leaching and gaseous emissions.[2] Phosphorus fertilization has also exceeded crop phosphorus use with only 10-20% of phosphorus fertilizer being utilized by crops.[6] As we lack tools capable of providing real-time information on the internal state of plants and fertilizer availability in the soil, we are unable to develop predictive models to guide farmers in effectively managing their fertilizer usage. Moreover, there is a need for significant investment in breeding strategies aimed at developing new varieties of plants with enhanced nitrogen, phosphorus and water use efficiency.

The complexity and global prevalence of agricultural systems can be addressed most effectively using transdisciplinary research. Where interdisciplinary research integrates multiple disciplines to address outstanding research questions, the NSF defines convergence research as the integration of disciplines to address a specific problem, typically of societal importance. Here, we define transdisciplinary research that involves convergent and interdisciplinary methods to co-create knowledge related to societal problems among stakeholders and researchers.[7,8] Transdisciplinary research is best suited to address this grand challenge since it has the most potential for creating knowledge and solutions that can be applied in social contexts, especially since its process relies on stakeholders and collaboration. In fact, transdisciplinary research has been used to address climate change challenges in regions around the world, including Europe, Africa, India, and Australia.[9–11] Transdisciplinary research should thus be applied to other problems, such as grand challenges in securing sustainable agricultural systems that meet the growing demands of the future.

Transdisciplinary research will be necessary to address global challenges and coordinate meaningful solutions in society. However, there is a need for more examples from biological and engineering applications of how transdisciplinary and convergence research can be implemented. Below, we present five case studies (Fig. 2) from two transdisciplinary and convergence research centers supported by the National Science Foundation as Science and Technology Centers (STCs), i.e., Science and Technologies for Phosphorus Sustainability (STEPS) Center, and Center for Research on Programmable Plant Systems (CROPPS), that contribute to our ongoing efforts in advancing sustainable and resilient agricultural systems. Collectively, our work spans across multiple scales. Starting on a whole-plant scale, we investigate how environmental inputs affect plant communication (signaling) and utilize synthetic biology approaches to harness this communication for reporting on resource availability. We then focus on the immediate plant environment (soil) and explore new fertilizer technologies to increase the availability of phosphorus in the soil. At a plant population scale, we develop tools for intimate interactions with plants to identify endophenotypes for improved breeding programs in the context of nitrogen and water use efficiency. Expanding to the ecosystem scale, we investigate phosphorus recovery from aquatic systems and develop water resource management strategies to mitigate the harmful effects of phosphorus leaching. Finally, we conduct research to understand stakeholder needs and priorities and incorporate these findings back into the Centers' research efforts. By showcasing our transdisciplinary endeavors, we hope to encourage the broad scientific community to join in this effort.

**Whole-plant reporting of below-ground resources for improved crop management**

Plants simultaneously exist in multiple distinct environments. For example, plant roots are embedded in terrestrial landscapes, and their shoots inhabit the air. To coordinate their physiology between these environments, plants have internal communication systems that enable whole-plant coordination and resource allocation.[12,13] These signaling communications exist across scales: within cells[14], between cells[15], from root-to-shoot[16], from leaf-to-leaf[17], and between plants and their environment[18]. For example, CLE 25 (CLAVATA3/EMBRYO-SURROUNDING REGION-RELATED 25 (CLE25) is a small peptide that moves through the xylem, from roots-to-shoots, to activate ABA-mediated drought stress responses in shoots.[19] Another peptide involved in long-distance signals is CEP1 (C-TERMINALLY ENCODED PEPTIDE 1) that moves from roots-to-shoots to activate nutrient foraging in response to low nitrogen.[20] MicroRNA2111 is a shoot-to-root small RNA signal that functions in balancing nodulation with rhizobial symbionts.[21] Plants also utilize rapid long-distance electrochemical (calcium ions and reactive oxygen species) and hydraulic signaling waves that are initiated in response to environmental stresses and then travel across the whole plant to induce stress resilience via systemic acquired acclimation.[22,23] By leveraging such diverse native long-distance signaling pathways, as shown in Fig. 3, we can gain access to information from inaccessible parts, such as the roots, via reporters in accessible parts of the plant, such as the plant canopy. Using approaches from plant science, synthetic biology and evolutionary genomics, we are developing reporter plants that illuminate the hidden communication between plants and their environment in real-time. By facilitating such whole-plant coordination, these reporter plants aim to transform our fundamental understanding of plant communication, and enable timely and efficient application of resources (e.g., fertilizers, water).

In addition to modifying native signaling networks to achieve these goals, researchers in the CROPPS are also building orthogonal sequences that can overcome the sometimes undesired off-target effects of simple switches based on native components. As such, synthetic approaches are being used in three major ways: 1) synthetic reporters that indicate the activity of native or orthogonal pathways, 2) synthetic switches that toggle native or orthogonal pathways for a desired effect, and 3) a combination of synthetic inputs and outputs that form a closed genetic circuit, which has been a successful use of synthetic biology tools in microorganisms and plants.[24–27] Together, such plant-compatible genetic circuits allow us to rebuild and engineer signaling pathways in the native plant host. Synthetic reporters can also be potentially multiplexed to allow a single plant to report on multiple environmental stresses or conditions, allowing for improved crop management strategies and assistance in molecular phenotype-based breeding efforts. Synthetic signal inputs can employ a combination of chemically induced genetic switches[28] and state-of-the-art optogenetic switches. Counterintuitively, the latter are usually derived from plant signaling proteins (e.g., photoreceptors) and were first used in in bacteria[29]. Only now are such optogenetic switches being introduced back into plants.[27] As an example, one of the first technologies developed within CROPPS combines newly developed genetic parts (including optogenetic switches) and long-distance signaling machinery, allowing for whole plant programming of below-ground reporting of nitrogen signals and responses.

In addition to engineering plants for below ground reporting, we are also exploring ways to harness the native microbiomes in the rhizosphere for improved crop management. Growing evidence points to the essential role of the rhizomicrobiome in facilitating plant nutrient uptake and acquisition by modulating soil nutrient availability, stimulating plant growth, and activating defense mechanisms.[30,31] Therefore, an improved understanding of plant–microbe-soil interactions, and the ability to control it by designing and leveraging heritability beneficial interactions could drastically reduce our dependence on external inputs such as synthetic fertilizers and increase resilience and sustainability. Current main-stream omics-centric approaches for investigating the rhizomicrobiome have led to rich phylogenetic and genomic information.[32] However, these approaches lack the resolution to understand the species-specific molecular underpinnings necessary to engineer the inputs/outputs due to the hyper complexity of the largely unculturable rhizomicrobiome, inaccessibility of the rhizosphere obscuring direct observation, and spatiotemporal heterogeneity in soils and gene expression.[33–35] Emerging single cell technologies, such as Single Cell Raman MicroSpectroscopy (SCRS), can begin to overcome these limitations in environmental systems due to the high resolution, *in situ,* non-destructive, and physical separation capabilities enabling downstream conventional "-omics" to provide for a multi-scale, molecular perspective of microbial systems.[36–40] Relying on single cell efforts to probe the rhizomicrobiome, we have made key strides in revealing the genotypes and phenotypes of the rhizomicrobiome communities associated with specific plant genotypes, as well as with plant root metabolites and exudates. For example, despite growing evidence for the important roles of nutrients and carbon accumulating organisms, such as polyphosphate accumulating organisms (PAOs), in the soil environment relevant to plant health, our understanding of the roles of PAOs and polyP in plant-microbe interactions is scarce.[41–44] For the first time, through cross-disciplinary collaboration, we are revealing the abundant PAO community in the rhizosphere, and identifying plant communication pathways that enrich PAOs.[45]

A challenge for modifying native plant signaling pathways in crop species is that most of our understanding of the networks that underlie signaling pathways comes from the small, weedy, genetic model organism *Arabidopsis thaliana*. Studies in *A. thaliana* have propelled our knowledge of plant genomes, development and function, but how can we translate these findings to crop species for which, in some cases, have a last shared a common ancestor with *A. thaliana* ~150 million years before present? To bridge this gap, we view each network or pathway through the lens of evolution (Fig. 3 - left). This workflow, in brief, requires 1) the inference of phylogeny for each component of the signaling transduction cascade / network, 2) reconciling the recovered gene trees with the accepted organismal relationships (termed gene tree - species tree reconciliation) and 3) integrating data on function and expression with the reconciled trees to provide insight on the extent of conservation (a form of evolutionary systems biology[46]). Importantly, this exercise permits both the identification of specific components of the signaling cascade to target for engineering (highly conserved components) and those that vary among organisms and thus represent opportunities to modify outputs when these components are introduced outside their native species. Using this approach, we are building our toolkit of evolutionary related signaling systems that can be deployed across species to achieve orthogonal signaling.

**Optimizing novel biomaterials for improved phosphorus fertilization**
An estimated 40% of the world's arable land is phosphorus limited.[47] Inorganic orthophosphate is available to plants but, in many soils, phosphate is either complexed with other molecules or is present in otherwise insoluble species, making most soil phosphorus largely biounavailable.[47,48] Thus, phosphorus fertilizers are necessary for agricultural productivity. Since crops have not been utilizing most of the applied fertilizers, it will be necessary to find strategies to either make soils less phosphate-limited or to more efficiently fertilize crops and reduce excess phosphorus leaching into the environment. Biomaterials, such as functionalized hydrogels and nanofertilizers are potential avenues to solve this problem, especially when co-created with biologists.

Functional hydrogels can be designed to cleave soluble, bioavailable phosphates from complexes. STEPS researchers have synthesized an organophosphate-degrading functionalized hydrogel by crosslinking poly(maleic anhydride-co-methyl vinyl ether) (PMAMVE) with ethylene diamine and reacting the remaining maleic anhydride units with with hydroxylamine.[49] This process forms hydroxamic acid functional groups within the PMAMVE gel. We evaluated the performance of the PMAMVE gels in decomposing dimethyl nitrophenyl phosphate (DMNP), an organophosphate molecule. The decomposition of DMNP in the maleic anhydride gels followed pseudo-first-order kinetics for all studied conditions. The spatial confinement of the hydroxamic acid groups inside the gel influenced the performance. The gels made of PMAMVE copolymers modified with hydroxamic acid offer a robust new system with high degradation efficiency, scalability and simple preparation. Current efforts toward creating materials that degrade organophosphates involve modifying the PMAMVE gels with deferoxamine (Desferal) and UiO-66 (Universitetet i Oslo) metal-organic framework. Once optimized by chemists and biologists in the STEPS Center, this hydrogel has the potential to improve phosphorus bioavailability of complexed phosphate that is already present in soils.

A second avenue of designing novel fertilization methods involves nanofertilizers. Per IUPAC (International Union of Pure and Applied Chemistry), a nanoparticle is classified as a particle with a size less than 100 nanometers in any one of the threedimensions.[50] Engineered nanoparticles often have new or novel physical or chemical properties compared to their bulk scale counterparts. There are essentially two main methods to synthesize them: top-down and bottom-up. The former involves grinding bulk material into the nanoparticle size while the latter synthesizes the material on the scale of nanometers using base or precursor materials. Top-down processes include different kinds of milling such as planetary or unitary ball mills. Bottom-up methods build from molecular or atomic clusters to nucleate and grow the desired nanoparticle. Bottom-up processes include chemical precipitation, sol-gel method co-precipitation, hydrothermal synthesis, microemulsion and electrochemical synthesis.[51] To further enhance the properties of nanoparticles, they are often surface modified or functionalized, a process in which inorganic and organic molecules are added to its surface, depending on the covalent or polar nature of the nanoparticle.[52,53]

In agriculture, functionalized nanoparticles are useful methods for obtaining controlled-release fertilizers. Nanoparticle controlled-release fertilizers exhibit better absorption and utilization of nutrients when compared to commercial fertilizers. Avila-Quezda *et al.* showed that nanofertilizers can reduce eutrophication due to loss of nutrients, exhibit a higher diffusion and solubility rate and have better controlled release properties when compared to commercially available controlled-release fertilizers.[54]

Recent work in the field has found that macronutrients such as N, K, P, Ca, Mg, and S can be combined with nanomaterials to enable the delivery of a specific quantity of nutrients to crops.[55] For example, Tarafdar *et al.* used phosphorus in tricalcium phosphate to synthesize fungal-mediated phosphorus nanoparticles.[56] This is an efficient way to maintain phosphorus in its usable form in the long-term as well as allow the mobilization of phosphorus for plants.[55]

Moreover, Liu and Lal showed that treating soybean seeds with carboxymethyl cellulose-stabilized hydroxyapatite nanoparticles by ultrasonic dispersion resulted in the most effective fertilizer among those tested, mainly because phosphorus demand is high at the local level and thus enhanced delivery at the nanoscale and in proximity facilitates proper absorption.[57] Interestingly, an important factor in nanoparticle controlled-release fertilizers is a uniform shape. When the particles are all consistent in both size and shape, the contact angle of the plant will not change when fertilizer is applied, ultimately leading to better absorption.[58]

In designing a nanofertilizer, a key parameter to consider is the fertilizer delivery method. If the application of the nanoparticles is foliar, then they are absorbed by trichomes, stomata, stigma, and hydathodes.[54] They are then transported through the phloem and xylem, where the nanoparticles enter the cells through endocytosis.[54,59] It is currently understood that a target size for nanofertilizers of about a few tens of nm can lead to enhanced absorption by the plant.[58] Nanofertilizers may also be designed to interact with or dissolve at the plant's outer surface, which requires consideration of a host of additional factors such as soil characteristics and plant cell wall chemistry.[59] Due to the physical, chemical and biological constraints of nanofertilizer development, materials scientists, biologists and social scientists within STEPS Center will optimize nanofertilizers to more efficiently and less wastefully deliver phosphate to plant cells. A wide array of factors must be considered, including economic, environmental and biochemical factors. For example, nanoparticle environmental and ecological persistence and impacts over short and long-term periods must be evaluated along with hazard assessments, risk analyses, predictions of large-scale economic benefits and costs and barriers to adoption in society[54].

Increased co-creation with the relevant stakeholders will address economic, regulatory and social barriers to functionalized nanoparticles and polymers in agriculture. Further collaborations between biologists, chemists, and engineers will also optimize these biomaterials for practical application in field settings. Rapid screening pipelines will aid in iterative testing, refining and co-creation of these biomaterials. 3D bioprinting has emerged as a technology that will enable rapid screening and testing of biomaterials such as hydrogels and nanoparticles. 3D bioprinting is an additive manufacturing technique in which protoplasts, or cells isolated by cell wall digestion, are extruded in a hydrogel with liquid media containing nutrients and hormones required to maintain cell viability for at least two weeks.[60] In plants, 3D bioprinting has been optimized by researchers within the STEPS for *Arabidopsis thaliana* and soybean embryonic, shoot or root cells.[61] Cells are extruded by the bioprinter reproducibly and at a high throughput, making it a suitable platform for screening cellular responses to each biomaterial (Fig. 4). Moreover, the 3D bioprinting techniques overcome current limitations of testing in protoplasts because protoplasts can be maintained for at least two weeks, allowing time for cell wall regrowth, cell division and longer-term studies than is typical.[59] By using 3D bioprinting to test cell responses to each material, it will be easier and faster for Center researchers to optimize and refine each material for their compatibility with plant cells and improved cellular phosphorus content before testing and applying them to crops in field sites. It will also be possible to better refine each material by testing it with a complex combination of factors, such as pH levels, matrix types and mineral compositions, that are closer approximations or models of field environments while still maintaining rapid, high throughput.

**Scalable field endophenotyping for high-throughput screening of plants**

To accelerate advancements in breeding new cultivars for increased yield, nutritional value, resource efficiency and resilience, it is necessary to develop technologies that enable high-throughput screening of plants and their associated organisms at molecular, tissue and whole-organism scales across time and space in the field. This process of measuring plant processes generates large datasets that have been useful in building genomic predictions for crop improvements.[62] Most plant phenotyping tasks have historically required manual effort, demanding a tedious and skilled workforce for completion.[63] Over the last decade, however, there has been a significant shift toward robotics and automation for high-throughput screening.[64] Efforts to estimate leaf area, color, shape, biomass, temperature, nutrient status and plant growth have seen the introduction of technologies such as ground-based robots, aerial drones and gantry systems. These systems are equipped with optical, hyperspectral and thermal imaging, as well as LIDAR and 3D scanning capabilities.[63–66] However, critical processes within the plant system remain inaccessible to interrogation and modulation for in-situ and in-field, and across diverse genetic backgrounds. For instance, current phenotyping efforts predominantly focus on visualization, leaving internal traits, or endophenotypes, largely unexplored. Moreover, while there is considerable focus on above-ground plant parts, below-ground aspects receive less attention due to technological constraints. Additionally, the existing robotic solutions are often rigid, whereas the nature of plant handling demands a delicate and gentle approach. To address these challenges, engineers and computer scientists in CROPPS are developing tools for in-field and high throughput interactions with above- and below-ground components of plant systems, as shown in Fig. 5.

Understanding water relations within plant tissue is necessary to enable improvement of crops function relative to resilience and sustainability of water use. Leaf water potential has been shown to impact vegetative growth and yield, particularly during extreme drought conditions, making it a promising trait for improving water-use efficiency in plants. Specifically, the water potential of the living tissue between the xylem and the stomata plays a critical role in plant function but has been largely inaccessible due to a lack of tools. Recently developed nano-reporter (AquaDust) enables in-situ and non-destructive measurement of water potential at this interface, providing unprecedented access to in-planta water status in a minimally invasive manner.[67] The nano-reporter is infiltrated into the leaves and transduces the local water potential into a fluorescent spectral signal. AquaDust has been successfully used in maize[68] and tomato[69], showing its potential as an endophenotyping method for crop development and has informed modeling efforts aimed at understanding water transport in plants. To enable automated and high-throughput endophenotyping using AquaDust, we are currently developing a soft-robotic system capable of automated leaf infiltration. The system consists of a soft gripper capable of autonomously infiltrating leaves with liquid solutions by gently squeezing them. In addition to enabling intimate interaction with the canopy, we are exploring approaches to access below-ground phenotyping, which faces significant challenges, primarily due to the sensitive nature of the plant rhizosphere.[70,71] Accessing this zone without disrupting the soil ecosystem proves difficult, with very few existing solutions for underground manipulation.[72] To address this, we are developing scalable underground soft robots capable of accessing roots and the rhizosphere with minimal damage, thereby establishing a direct connection with roots and the microbiome.

**Phosphorus recovery from aquatic ecosystems**

Phosphorus can be present in many different forms (e.g., inorganic/organic, soluble/particulate) that dictate its behavior in water and soil matrices. In soils, many phosphorus species are present either as phosphate complexed with iron and other metal ions or are contained within organic molecules often referred to as legacy phosphorus.[48,73,74] Similarly, phosphorus is present as different species in aquatic systems. The soluble, reactive phosphorus fraction (orthophosphate) is directly available for biological uptake. Orthophosphate is also more readily recoverable using phosphorus treatment technologies applied in water and wastewater treatment systems, whereas the non-reactive phosphorus fraction is less studied. However, consideration of this fraction, which can comprise the majority of phosphorus in some water matrices[75,76], is important for sustainable phosphorus management. Collaborations between biologists and engineers within the STEPS Center can advance efforts to manage and utilize these 'lost' phosphorus fractions more efficiently. For example, different physical, chemical and biological approaches can be used to facilitate phosphorus transformations to the target form.[75]

In nature, phosphorus transformations occur at varying rates. Engineered systems can accelerate such processes to facilitate efficient phosphorus removal, recovery and reuse in agriculture. For example, wastewater treatment facilities create conditions to remove phosphorus either chemically or biologically. Chemical transformation often involves adding iron or aluminum salts to precipitate phosphorus, where soluble phosphate (e.g., $PO_4^{3-}$) is transformed into a solid form (e.g., $AlPO_{4(s)}$). Phosphorus transformation can also derive from physical-chemical processes, including oxidation processes. Redox reactions involve the transfer of electrons between chemical species. In water, phosphorus is typically present in the +5 valence state such that phosphorus bonds are not specifically targeted by oxidation. However, when phosphorus is incorporated in complex organic material, advanced oxidation processes may facilitate phosphorus recovery by partially oxidizing the phosphorus-containing materials into molecules more susceptible to hydrolytic conversion to phosphate or completely oxidizing the organic to yield phosphate.[75] A wide array of advanced oxidation processes have been tested as a means of breaking down complex or recalcitrant organics in water. These processes feature the generation of powerful oxidizing radical species such as hydroxyl radicals (OH·) capable of mineralizing organics to simple components such as $PO_4^{3-}$, water and carbon dioxide.

Electrochemical processes are a type of advanced oxidation processes that use direct current to generate chemical reactions in-situ by transferring electrons from an anode to a cathode (Fig. 6a). This negates the need for transportation and storage of auxiliary chemicals. In electrooxidation, non-active electrode materials such as boron-doped diamond or $Ti_4O_7$ are often used to yield oxidation processes via direct oxidation on the electrode surface or indirect oxidation by hydroxyl radicals or other oxidants generated in-situ.[77] Mallick et al. observed that electrooxidation featuring direct electron transfer was able to transform soluble non-reactive phosphorus to reactive forms, although the presence of organics in the water matrix interfered with process efficiency.[76] The soluble reactive phosphorus compounds containing organic phosphoester P-O-C bonds were more susceptible to oxidation compared to inorganic anhydride P-O-P bonds.[78] Notably, complete conversion of non-reactive phosphorus to orthophosphate required high electrooxidation energy inputs such that incomplete conversion, when advanced oxidation is already in place as part of the water treatment regime, may be the most relevant means of implementation.

An example of biological transformation involves Polyphosphate-Accumulating Organisms (PAOs) in a process known as enhanced biological phosphorus removal (EBPR). PAOs take up soluble phosphate and transform it into intracellular chains of phosphate that are called polyphosphate (poly-P) (Fig. 6b). PAOs are unique because they can hyperaccumulate phosphorus to levels exceeding 15% of a cell's total dry biomass. In water resource recovery facilities (WRRFs), the most common PAOs are *Candidatus*

*Accumulibacter*, *Candidatus Phosphoribacter* (previously *Tetrasphaera*), and *Dechloromonas*.[79] EBPR depends on cycling between anaerobic and aerobic conditions to enrich PAOs. Under anaerobic conditions, PAOs consume carbon substrates such as volatile fatty acids, glucose or amino acids and store them as polyhydroxyalkanoates (PHA), glycogen or free amino acids.[80–83] Polyphosphate is a source of energy used to synthesize these carbon storage molecules, and its catabolism results in release of inorganic orthophosphate (Pi) from the cells. Under aerobic conditions, stored carbon is consumed to produce energy for cell growth and reproduction. In the process, polyphosphate stores are replenished, resulting in a net removal of phosphorus from the wastewater. After settling, collecting, and treating the microbial biomass (called biosolids in the wastewater industry) that is rich in phosphorus, it is often applied to agricultural lands. The forms of phosphorus, such as poly-P, are more bioavailable to the soil and plants than is chemically precipitated phosphorus. The intracellular poly-P is slowly degraded and released to the soil and plants (Fig. 6b). Digesting the settled biomass in anaerobic digesters releases the poly-P as orthophosphate, which can then be converted into field-ready fertilizers such as struvite at WRRFs. To the best of our knowledge, outside of the wastewater treatment field, there are no other applications of PAOs to remove phosphorus from aqueous matrices.

PAOs may play a role in P transformations in natural environments, including agricultural environments and those adjacent to agricultural settings.[41] Biologically-derived poly-P is ubiquitous in nature, though the contribution of specific PAOs in P cycling in natural environments is not well understood.[41,84] However, there is evidence of PAO-mediated phosphorus cycling in these environments. Saia et al. conducted *ex-situ* experiments on native stream biofilms, demonstrating characteristic phosphorus uptake and release capabilities by these microbial communities.[85] Coupled with evidence of intracellularly-stored poly-P, this study demonstrated that poly-P accumulators in freshwater environments may contribute to phosphorus cycling, particularly during diel oxygenic conditions. Furthermore, Taylor et al. demonstrated that stream biofilms in nutrient-depleted sites stored more intracellular poly-P than those in nutrient-rich sites.[86] This study suggested that microbial communities in environments with higher legacy phosphorus store less poly-P. PAOs identified in EBPR have also been identified in sediments. Watson et al. detected *Ca. Accumulibacter* in Columbia River estuary sediments that exhibited an increase in poly-P during the day.[87] Understanding the roles of PAOs for phosphorus transformations in natural environments and specifically how PAOs contribute to phosphorus cycling in agricultural soils is an opportunity for future research. By advancing understanding of both biological and physical-chemical phosphorus transformation processes, biologists and engineers can work together to better harness these powerful systems to improve phosphorus management by transforming phosphorus into forms that are more conducive to removal (to enhance protection of environmental waters from eutrophication), recovery (to encourage a circular phosphorus economy), and reuse (to support global food production).

**Integrating stakeholder perceptions and needs into research and innovation for phosphorus sustainability**

Developing sustainable solutions for phosphorus management requires new technological innovations and improved management practices, as well as a deeper understanding and integration of stakeholder needs and priorities. By integrating stakeholder needs and priorities within research and innovation cycles, the proportion of technologies, management practices and innovations (e.g. new crop varieties, smart fertilizers, removal/recovery systems) that will ultimately be adopted by end-users increases substantially. In addition, better understanding and incorporating stakeholder needs and priorities within early phases of research and innovation align with broader concepts of responsible innovation, that seek to better align scientific research and innovation efforts with societal needs and wants.[88,89]

For these aforementioned reasons, the STEPS Center conducts research to identify the perceptions and needs of diverse stakeholders involved in phosphorus management and sustainability using a range of approaches. Among other examples, a survey was conducted to understand the perceptions, needs, and potential concerns of stakeholders for achieving phosphorus sustainability.[89] Building on the survey results of the survey, a targeted set of interviews were conducted with key stakeholders involved in a subset of STEPS-funded projects that aimed to develop new technologies or strategies to improve sustainability of phosphorus management in specific contexts.[90]

The research team developed the survey to investigate stakeholder views and needs for achieving phosphorus sustainability using the online platform Qualtrics. The survey was composed of 14 multiple-choice and open-ended questions, and was deployed to expert stakeholders in the U.S. and abroad in the fall of 2022 using email and posted on social media (LinkedIn) through the Sustainable Phosphorus Alliance, which is a STEPS partner. Based on responses of 96 survey participants, stakeholders largely considered phosphorus management to be unsustainable and they were concerned about our current ability to manage phosphorus in a sustainable way.[89] These responses were based on the fact that phosphorus is mined from nonrenewable phosphate rock, there are significant challenges to phosphorus recovery and reuse, phosphorus management in agriculture is inefficient, and the environmental impacts associated with the inefficient management of phosphorus. This study also found that stakeholders reported a range of needs to achieve phosphorus sustainability, including better management practices, new technologies, new or better regulations, and ways in which to engage other stakeholders. While these findings are not entirely surprising for researchers and practitioners working in phosphorus sustainability fields, this study was the first to document stakeholder needs and priorities to achieve phosphorus sustainability using a structured and rigorous approach based in social sciences.[89]

Next, the research team identified three case studies that represented promising technologies or new strategies to improve phosphorus sustainability across the STEPS Center to investigate stakeholder perceptions and needs in these contexts more deeply. The case studies selected were 1) urine diversion in buildings, 2) EBPR in wastewater treatment, and 3) isotope tracking in natural waters. A total of 37 interviews were conducted with key stakeholders involved in these case studies in the spring of 2023 to better understand their perceptions, needs, and potential concerns for achieving phosphorus sustainability.[90] Key findings from this work include the fact that stakeholders identified a common set of economic, regulatory and societal issues as main factors that affect the adoption of new scientific or technical solutions for phosphorus recovery. For example, the cost of water and phosphate rock were identified as primary drivers to the adoption of urine diversion in buildings, whereas technical improvements were a primary driver in the EBPR case. At the same time, barriers to technology adoption included technical, regulatory and economic challenges, such as societal push-back associated with negative perceptions of urine diverting toilets in the urine case or technical challenges to EBPR. Findings from this work are being integrated back into STEPS-funded research projects to improve not just technical aspects but also engagement efforts associated with these cases.[91] Following these efforts, the research team is also developing collaborative working groups within specific STEPS projects to co-create solutions with stakeholders in specific contexts and settings, as well as developing best practices and guidance to follow for other researchers who aim to engage stakeholders in their work related to phosphorus sustainability (Grieger et al. - in preparation).

Overall, these examples highlight how trans- and convergence research can be conducted in specific projects focused on an aspect of phosphorus sustainability, through the integration of natural sciences, environmental engineering and social sciences.

**Integrating stakeholder perceptions for nitrogen and water management**
The CROPPS center integrates stakeholder perspectives in the development of scalable field computing technologies intended to support nitrogen and water management. In addition to typical external engagements via surveys and other instruments, the center looks inward to the trajectory of its own innovations to ensure that they match day-to-day realities of rural environments where the technologies are imagined to be eventually deployed. Looking both *early* and *inward* in transdisciplinary research involving computing is important because prior literature indicates that computer systems developers inadvertently embed their own values into computing technologies - a term widely known as values in design.[92–94] As a result, the team conducted a research study with two components:
- A 24-month autobiographical study where researchers deployed experimental networking technologies for scalable field computing while critically reflecting on their day-to-day systems building to anticipate potential societal impacts of early technical design decisions
- A set of seven semi-structured interviews with researchers and farmers involved in early phase adoption and adaptation of similar networked technologies for data-driven agriculture to understand their challenges, attitudes, and visions of the emerging technologies

Key findings from the study include a one-hundred year gap between the *seamless* visions of researchers and the *seamful* realities under which the technologies are being built and expected to operate.[95] For example, one participant described realities where "some of my neighbors farm like it is 1950". Furthermore, as a result of the environments under which research is conducted (e.g., ample university IT support and research funding), researchers maintain faith in an *eventual seamlessness* of the technologies in the contexts where they will be deployed (i.e., rural farms).

The century gap is crucial to acknowledge because it could affect consumer adoption of technologies built in the lab to address global challenges, but not grounded in the day-to-day realities of farms. More importantly, the findings raised three implications for the ongoing design, implementation, and evaluation of scalable field computing technologies within CROPPS:
- Supporting the right to tinker with the technologies to fit the particular environment where they could be deployed
- Documenting invisible work (e.g., hardware and cloud computing service choices) and their effects on the design decisions
- Bridge the gulf to farms by recognizing farmers as natural tinkerers that have much insights to offer to technology designs and implementation

As a result, the third implication has spurred further *external* research studies in CROPPS to guide internal vanguard research visions. Moreover, these findings formed part of a foundation for center-wide workshops to engage the center's researchers in imagining alternative visions to the current trajectory of agricultural production and technological design.

**Conclusion**

With a changing climate, a growing population, and decreasing availability of resources such as water, phosphorus and nitrogen, there is an urgent need to improve the productivity and sustainability of crop-based agriculture. Transdisciplinary research has emerged as a research model that can address the coupled scientific, technological and societal dimensions of this challenge in a manner that defines successful convergence research. In this review, we highlight several case studies concerning the implementation of transdisciplinary and convergence research, including research to develop reporter plants of below-ground nutrient and water availability as well as integration of plant-microbial interactions as efficient strategies for agricultural resource management. Moreover, plant-informed early optimization of new phosphorus fertilization technologies is in development to help reduce the overuse of applied phosphorus fertilizer by either improving soil phosphorus bioavailability or implementing more efficient nanofertilizer methods. Integrating nanotechnology with robotics, we show how we can accelerate advancements in breeding new cultivars. Finally, phosphorus recovery strategies are being developed using electrooxidation or polyphosphate-accumulating organism systems to help resolve the problem of phosphorus contamination of aquatic ecosystems. Each center has also identified stakeholder needs and barriers to adoption of research projects to help guide their respective institutions towards advancing work that can successfully advance their target societal goals.

As transdisciplinary research is implemented in future applications, it will be important to consider a number of factors such as diversity in collaborations, cross-training among disciplines, and co-creation with stakeholders.[7,96] Transdisciplinary convergence research efficiently facilitates innovation as each team can combine their expertise in new ways that, in isolation, they would not be able to implement in a timely manner, if at all. For example, endophenotyping enabled by nanoreporters and robotics relies on collaboration between engineers and plant biologists, while fertilizer biomaterial optimization relies on collaboration between chemists, materials scientists and plant biologists. Both of these developments provide new opportunities to engage stakeholders in the design of application-appropriate research questions and technologies. Seeking out opportunities to co-create in multidisciplinary teams will thus help bring innovative ideas to fruition without the need for an impossible task of one individual team building expertise in all disciplines. Furthermore, innovation and technology does not exist in a vacuum. Social factors such as stakeholder needs and perceptions, governmental regulations, economic factors, environmental risks and practicality of using the technology all need to be considered.[7,8,97]. To do this, stakeholder engagement and co-creation are key to guiding the focus of transdisciplinary research projects in real time. Accordingly, collaboration with researchers within the social sciences is necessary to produce meaningful innovations. Thus, transdisciplinary research implemented to its full potential brings efficiency and direction to research and innovation towards solving complex global challenges.

**Conflict of interests**

The authors declare no conflicts of interest.

**Acknowledgements**

This work was supported by the U.S. National Science Foundation (NSF) under awards CBET-2019435 (STEPS) and DBI-2019674 (CROPPS). V.B. is supported by the Schmidt Science Fellows program and the Kavli Institute at Cornell (KIC) Postdoctoral Fellowship. I.M. is supported by the NSF Postdoctoral Research Fellowships in Plant Biology Program through award number IOS-2305774.

**Figures**

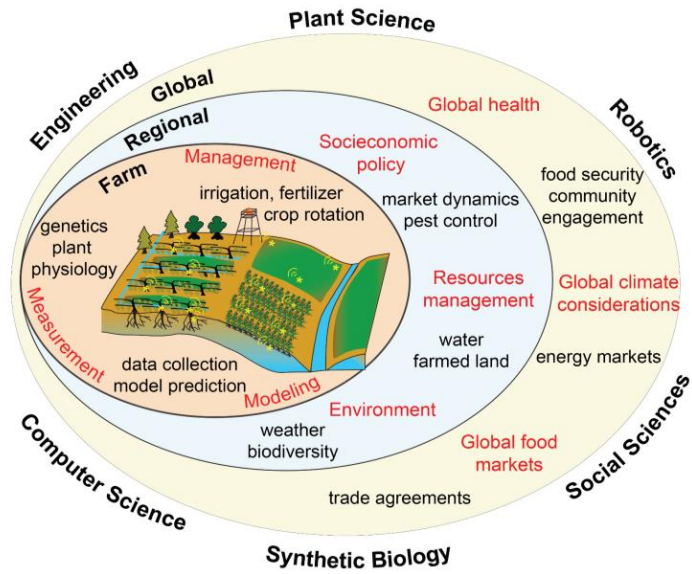

**Fig. 1.** Multiscale management of agricultural systems from farms to regional and global scales. Changing agricultural practices for a sustainable and resilient future requires collaboration across plant science, engineering, and social sciences, involving all stakeholders, including farmers and consumers.

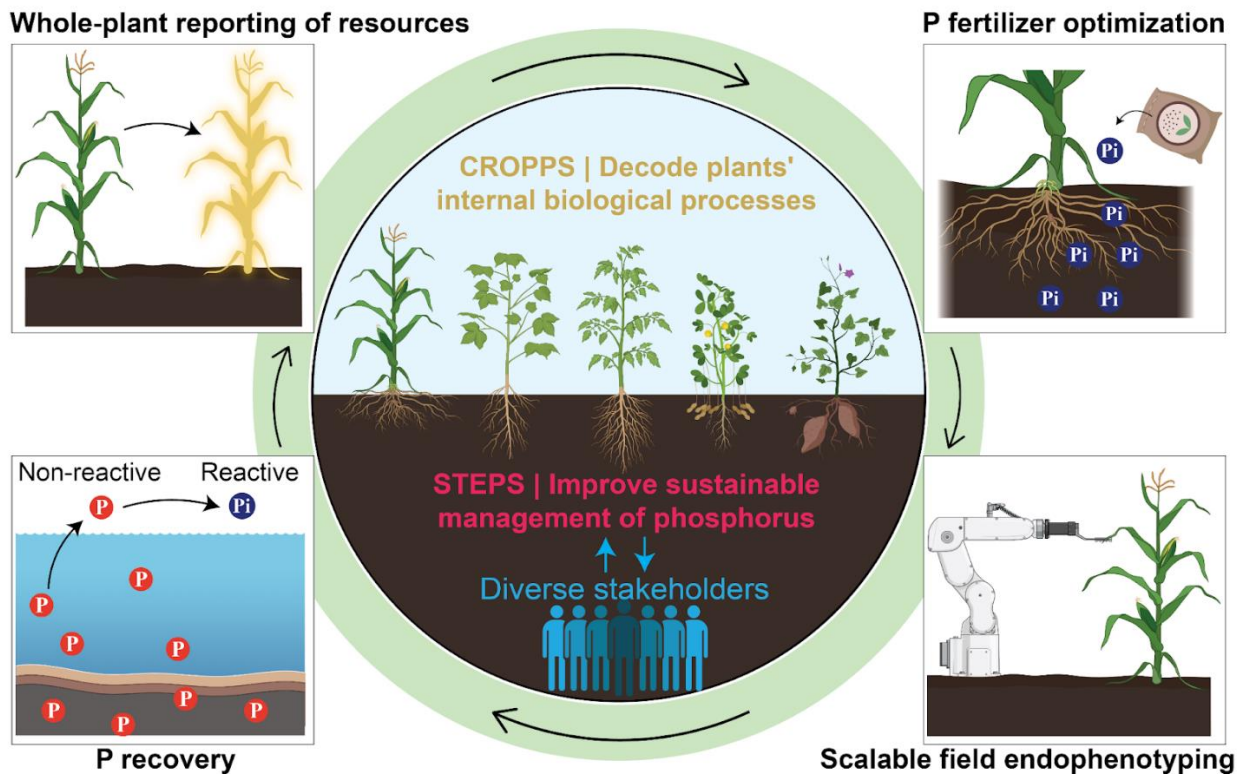

**Fig. 2.** Case studies of transdisciplinary collaborations for advancing sustainable and resilient agricultural systems including whole-plant reporting of environmental inputs via plant communication, new fertilizer technologies for increased phosphorus availability, field tools for breeding programs, phosphorus recovery from aquatic systems, and engagement with diverse stakeholders. These case studies are being pursued in two NSF Science and Technologies Centers, the Science and Technologies for Phosphorus Sustainability (STEPS) Center, and Center for Research on Programmable Plant Systems (CROPPS).

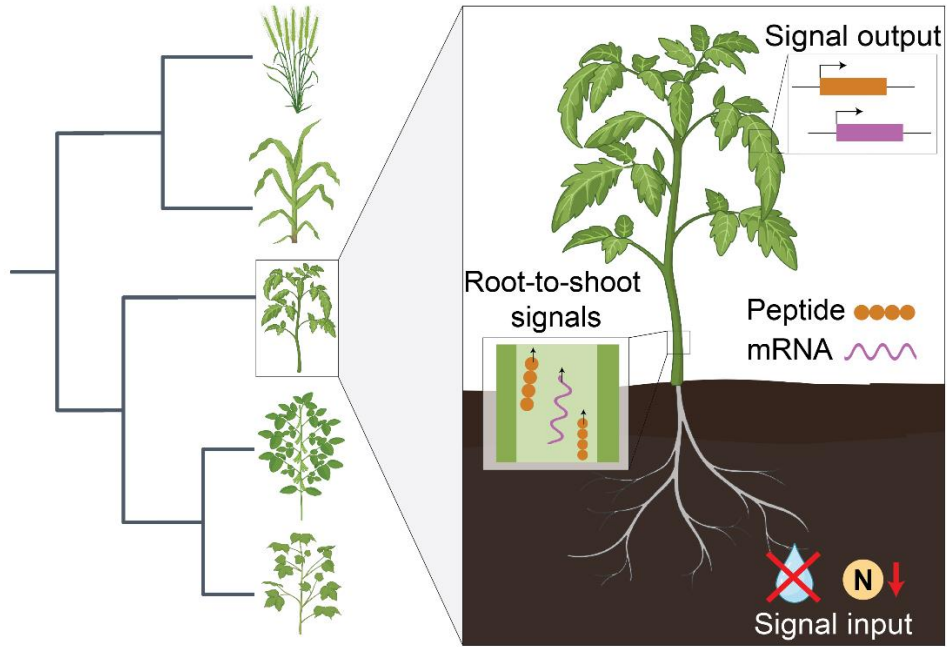

**Fig. 3.** We bioengineer plants to report on below-ground resources (e.g., water and nitrogen availability) by leveraging their native long-distance signaling pathways and use evolutionary genomics to translate this technology into different species and genetic backgrounds relevant for farmers.

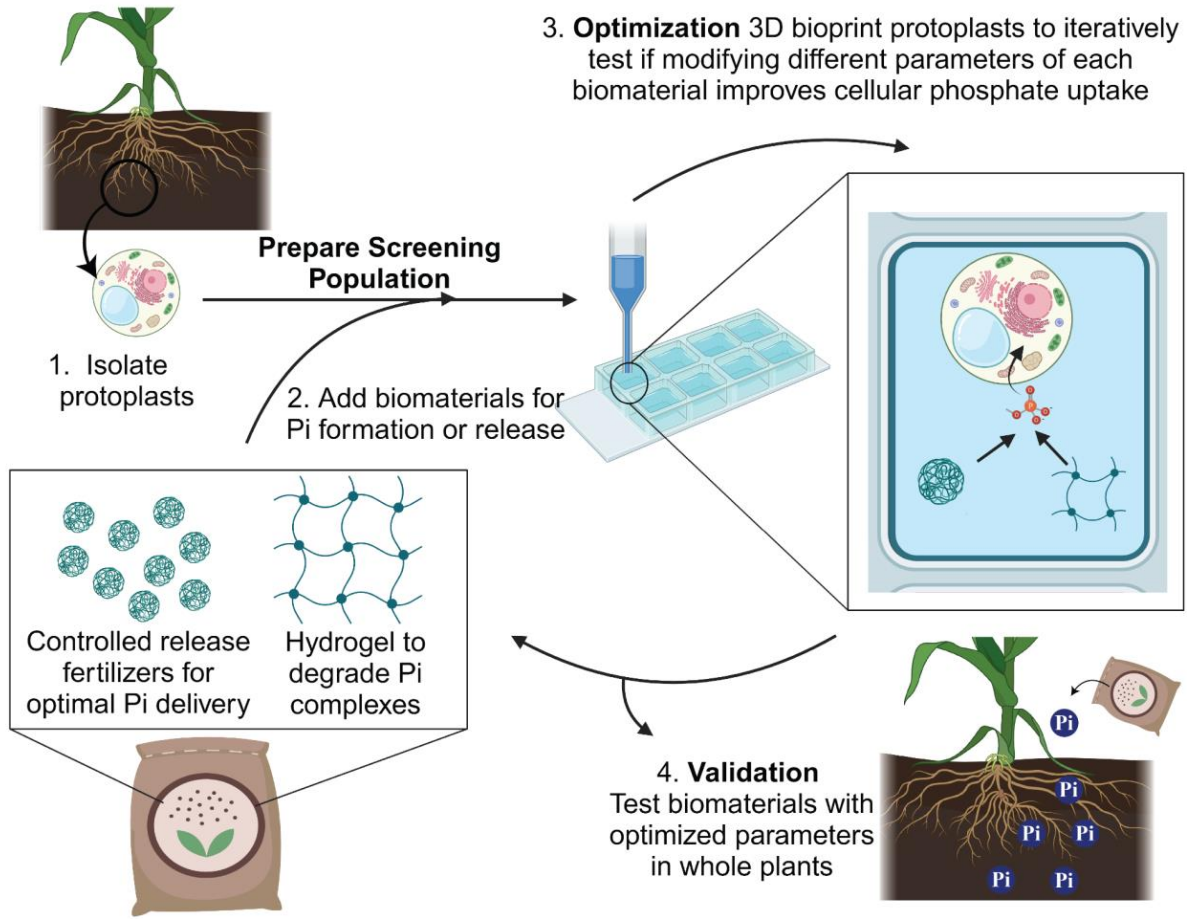

**Fig. 4.** Improve phosphate (Pi)-based fertilizer chemistry prototyping using a 3D bioprinting approach.

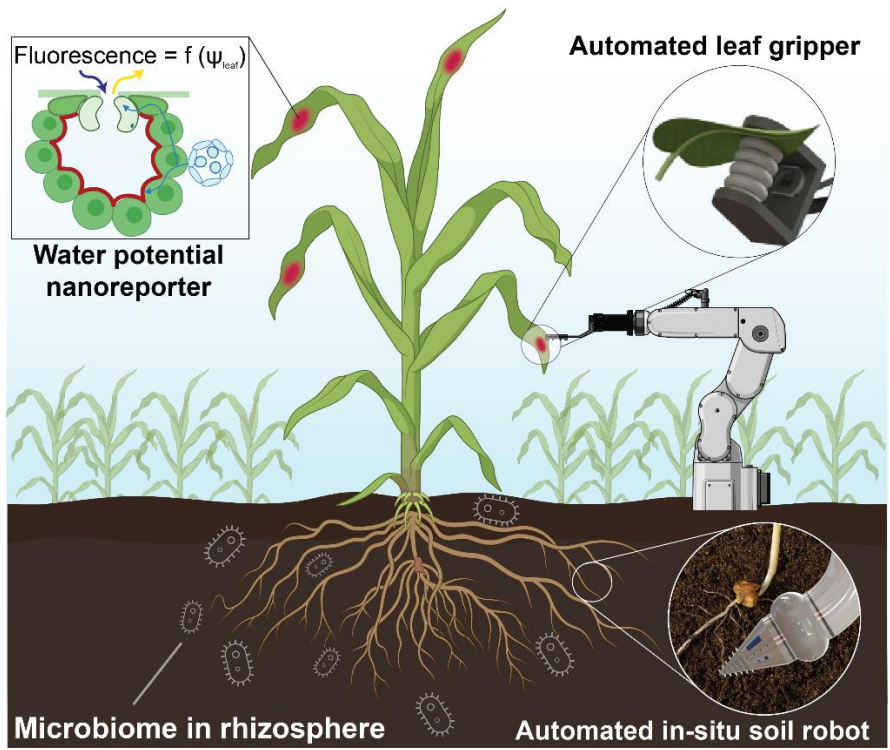

**Fig. 5.** Automated soft robotic leaf gripper for measurement of water status in plants using a water potential nano-reporter infiltrated in the leaves, and a soil robot for automated and non-invasive below-ground phenotyping of soil microbiome and root structure.

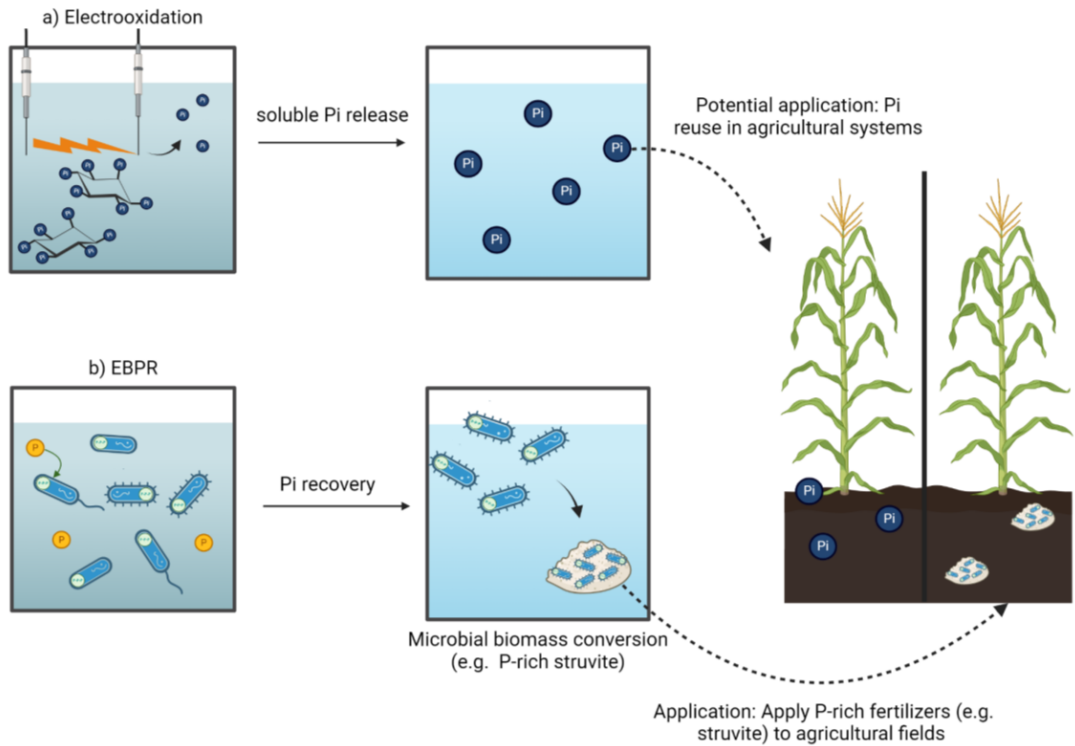

**Fig. 6.** Phosphorus transformations to facilitate removal and recovery of phosphorus from wastewater using: a) Electrooxidation to release soluble phosphorus (Pi), and b) Enhanced biological phosphorus removal (EBPR) to store phosphorus (P) in polyphosphate chains as well as potential downstream applications of reusing recovered phosphorus.